\documentclass[useAMS,usenatbib]{mn2e}
\usepackage[dvips]{graphicx}
\usepackage{textcomp}
\title[Near Infrared Angular Diameters of a few AGBs]{Near Infrared Angular diameters of a few AGB variables by Lunar Occultations}
\author[Tapas Baug and T. Chandrasekhar]{Tapas Baug\thanks{E-mail: tapasb@prl.res.in (TB); chandra@prl.res.in (TC)}
 and T. Chandrasekhar\footnotemark[1]\thanks{This file has been amended to highlight the proper use of \LaTeXe\ code
 with the class file.}\\Physical Research Laboratory, Ahmedabad 380009, India\\}
\begin{document}

\date{Accepted}

\pagerange{\pageref{firstpage}--\pageref{lastpage}} \pubyear{2011}

\maketitle

\label{firstpage}
\begin{abstract}
The uniform disk (UD) angular diameter measurements of two oxygen-rich Mira variables (AW Aur and BS Aur) and three semiregular (SRb) variables (GP Tau,
 RS Cap, RT Cap), in near Infrared $K$-band (2.2 $\mu m$) by lunar occultation observations are reported. UD angular diameters of the two Miras and one
 SRV are first time measurements. In addition a method of predicting angular diameters from (V-K) colour is discussed and applied to the five sources.
 The effect of mass-loss enhancing measured $K$-band diameters is examined for Miras using ($K-[12]$) colour excess as an index. In our sample the
 measured angular diameter of one of the Miras (BS Aur) is found enhanced by nearly 40\% compared to its expected value, possibly due to mass loss effects
 leading to formation of a circumstellar shell.
\end{abstract}


\section{Introduction}
Asymptotic giant branch (AGB) stars are highly evolved cool stars in the last stages of their stellar evolution before turning into planetary nebulae. The
 high mass loss rates ($10^{-7}-10^{-6}$ $M_\odot yr^{-1}$) and relatively low surface temperatures of these stars provide a habitable zone for several
 molecules like $TiO$, $VO$, $H_2O$ and $CO$ in their extended atmospheres. A very common characteristic of AGB stars is the long period variability of
 their radiative output, mainly, due to pulsation of their atmospheres though episodic ejection of dust can also contribute to the variability (Lattanzio
 \& Wood, 2003). Traditionally AGB stars have been classified according to their visual variability amplitude in magnitude as (1) classical Mira variables
 (visual amplitude $>$ 2.5), clearly defined periodicity in the range 100-1000 days, (2) Semi-regular variables (SRV): $SRa$ - visual amplitude $<$ 2.5 and
 periods in the range 35-1200 days; $SRb$ - visual amplitude $<$ 2.5 with poorly defined periods, (3) Irregular variables with small amplitude and no
 definite periods. In addition there is also a class of Super-giant semi-regular variables ($SRc$) . Micro-lensing surveys such as $MACHO$ (Alcock et al.,
 1995), $EROS$ (Aubourg et al., 1993) and $OGLE$ (Udalski et al., 1993) during the last 15 years have produced many high quality light curves of the AGB
 stars upto $V$-mag $\sim 20$ which have discerned distinct periodicities in objects hitherto  classified as irregular variables.

A recent important class of AGB variables too faint to be found in GCVS (General Catalog of Variable Stars; Samus et. al., 2009) consist of dust enshrouded
 infrared variables, found in Infrared surveys, which pulsate with larger amplitude in K-band (2.2 $\mu m$) ($K$-amp $\sim$ 3) and longer period ($\geq$ 600
 days) than optical Mira Variables (Whitelock et al., 1994; Wood, Habing \& McGregor, 1998). The dust enshrouded IR variables are considered to be in a more advanced state
 of evolution than classical Miras.

\begin{table*}
 \centering
 \begin{minipage}{160mm}
  \caption{Observational details of $Lunar$ $Occultation$ events.}
  \begin{tabular}{@{}ccccccccccc@{}}
  \hline
  Source & JD Obs.  &   Phot. & Lunar&   Alt.   &   Detector    &Samp. time& S/N &$V_{comp}^\ddag$&  PA$^\S$  & CA$^\natural$   \\
         & 2450000+ &Phase$^\dag$& Phase&($^\circ$)&     used      &(millisec)&     &(m/ms)&($^\circ$)&($^\circ$)\\
         &          &         &(days)&          &               &          &     &          &          &         \\

  \hline 
  AW Aur & 3802.082 & 0.95  &  7.6 &   85.5   &InSb Photometer&   1.00   & 14  &  0.616   &   71.3   &   16.4   \\
  BS Aur & 3775.306 & 0.83  & 10.2 &   45.1   &InSb Photometer&   1.00   & 20  &  0.807   &   87.0   &   13.9   \\
  GP Tau & 2710.251 & 0.43  &  8.6 &   29.5   &InSb Photometer&   2.00   & 19  &  0.486   &  148.1   &   55.2   \\
  RS Cap & 4804.023 & 0.43  &  5.8 &   36.0   &HgCdTe subarray&   7.29   & 39  &  0.339   &   -5.7   &   58.2   \\
  RT Cap & 4802.908 & 0.72  &  4.7 &   36.9   &HgCdTe subarray&   8.75   & 48  &  0.516   &   91.0   &   27.5   \\
\hline 
\end{tabular}
\begin{flushleft}
 $^\dag$ Photometric phase is derived for the epoch of our observation from GCVS or ASAS catalogues. It varies from 0 to 1 with 0.5 signifying minimum
 light.\\ $\ddag$ $V_{comp}$ refers to the predicted velocity component of the moon in the direction of occultation.\\$\S$ PA is the position angle of the
 point of occultation on the lunar limb measured from North to East.\\ $\natural$ CA is the contact angle between the direction of lunar velocity and the
 direction of occultation.
\end{flushleft}
\end{minipage}
\end{table*}

High angular measurements of angular sizes of Mira variables at different phases of their pulsation cycle provide an important direct means of
 understanding their atmospheric extension and pulsation properties. However, large opacities due to absorption by molecular species in their atmospheres
 mask the dominant continuum radiation from the photosphere. Consequently, photospheric angular size measurements are affected differently in different
 filter bands which has been known for sometime (Haniff, Scholz \& Tuthill, 1995). In recent years there have been many high quality measurements of Mira
 variables at $IR$ wavelengths but measurements of SRVs are relatively few. Mennesson et al. (2002) found $L^{\prime}$-band diameters of several oxygen
 rich Miras were much larger (25\% to 100\%) than those measured in the broad $K$-band and ascribed it to a wavelength dependent transparency of an optically
 thin gaseous shell around the star. In a multi-epoch interferometric study of two Mira variables spread over several pulsation cycles, Thompson,
 Creech-Eakman \& van Belle (2002) report variations in narrow band angular sizes within the $K$-band (2.0 - 2.4 $\mu m$) and attribute them to molecular
 absorptions. Perrin et al. (2004) observed several Miras in the narrow bands around 2.2 $\mu m$ and found systemically larger diameters in bands
 contaminated by water vapour or CO. Millan-Gabet et al. (2005) report from simultaneous measurements in $J$, $H$ and $K^{\prime}$-bands of 23 Miras, a
 systematic increase of angular size with wavelength (25\%) from $J$ to $H$ to $K^{\prime}$. Mondal \& Chandrasekhar (2005) find a 20\% increase in their
 $K$-band Lunar Occultation (LO) measurement of one Mira (U Ari) compared to a reported $H$-band measurement at the same phase. These authors also reported
 that two SR variables do not show any phase variation in their $K$ and $L^{\prime}$ angular diameters. In a detailed 3 telescope interferometric study in
 the $H$-band Ragland et al. (2006) find that almost all Miras show an asymmetry in their brightness distribution and attribute it to the formation of an
 inhomogeneous translucent molecular screen located at about 1.5 to 2.5 stellar radii. Eisner et al. (2007) using higher spectral resolution interferometric
 observations of a Mira (R Vir) find the measured radius of emission varies substantially from 2.0 to 2.4 $\mu m$. They infer that most of the molecular
 opacity arises predominantly due to $H_2O$ at about twice the stellar photospheric radius. Propagating shocks associated with Mira pulsation provide a
 mechanism for lifting the molecular layer to the observed location. Woodruff et al. (2009) in a spectro-interferometric study of 3 Miras from 1.1 to 3.8
 $\mu m$ report strong size variations with wavelength probing zones of $H_2O$, $CO$, $OH$ and dust. The variation in UD angular diameters by a factor of
 two from 1.0 to 3.0 $\mu m$ consolidates the picture of a Mira atmosphere consisting of molecular shells and time dependent densities and temperatures.


In this paper we present high angular resolution measurements in the broad $K$-band using the Lunar Occultation technique of two oxygen rich Miras and
 three semiregular variables. Uniform Disk (UD) angular diameters are reported. For the two Miras and one SRV (GP Tau) UD diameters are reported for the
 first time.

\begin{table*}
 \centering
 \begin{minipage}{140mm}
  \caption{Observed Source details}
  \begin{tabular}{@{}lccccrrccc@{}}
  \hline
  \multicolumn{2}{|c|}{Source}&Spectral& Variability &    $K$-mag    &\multicolumn{2}{|c|}{$V$-mag}& ($K-[12]$)  & Period    & Ref.\\
     IRC No.     &     Name   &  Type  &     Type    &               &    max     &      min       &colour excess& (days)    & cat.\\
\hline 
          +30123 &   AW Aur   & M5-M9  &     Mira    & 2.34$\pm$0.07 &  10.10     &    17.10       &    1.1     &445$\pm$10 & GCVS \\
          +30136 &   BS Aur   & M8-M9  &     Mira    & 2.15$\pm$0.06 &  10.20     &   $>$15.00     &    2.6     &467$\pm$05 & GCVS\\
          +20116 &   GP Tau   & M7     &     SRb     & 0.29$\pm$0.07 &   9.57     &    10.40       &    0.9     &109$\pm$10 & ASAS\\
 \textminus20596 &   RS Cap   & M6/M7  &     SRb     &-0.39$\pm$0.04 &   7.90     &     8.36       &    1.2     &193$\pm$15 & ASAS\\
 \textminus20585 &   RT Cap   & C      &     SRb     & 0.55$\pm$0.06 &   7.66     &     8.61       &    0.2     &389$\pm$10 & ASAS\\
\hline 
\end{tabular} 
\end{minipage}
\end{table*}

\section[]{Observations and Data Analysis}
The LO observations of all but one of these sources were carried out in the near-Infrared broad-$K$-band (2.2 $\mu m$/0.4 $\mu m$). The bright source RS
 Cap was observed in narrow $CO$-band filter (2.37$\mu m$/0.1 $\mu m$) to avoid saturation effects. The details of the observations are listed in Table.1.
 The 1.2 m telescope of Mt Abu IR-observatory (lat: 24$^\circ$ 39$^\prime$ 10$^{\prime\prime}$ N, Long: 72$^\circ$ 46$^\prime$ 47$^{\prime\prime}$ E, Alt:
 1680m ) was used for observations with two different IR detector systems. All are disappearance events at lunar phase measured in days after new moon as
 listed in Table.1. Three sources (AW Aur, BS Aur and GP Tau) were observed with an older system using a single element InSb detector with an effective
 field of view about 10 arcsec in the sky. The details of this system can be found in Chandrasekhar (2005). The other two sources were recorded using the
 10 X 10 pixel ($5'' X 5''$) subarray of a 256 X 256 pixel Mercury Cadmium Telluride (MCT) detector array of the NICMOS IR-camera. Details of the subarray
 mode of operation for lunar occultations and analysis are extensively discussed in a recent paper (Chandrasekhar and Baug, 2010). Typically an observing
 run in the subarray mode consists of initiating the data acquisition procedure for recording 4800 sub frames about 20 seconds before the predicted time
 of the event. About 15 seconds after the predicted event time the telescope is rapidly switched to nearby sky to record sky frames. Sky subtracted
 sub-array frames are used to derive the light curve.

The light curve is carefully analysed to extract the uniform disk (UD) angular diameter of the stellar source using the method of non-linear least squares
 (NLS) first enunciated by Nather and McCants (1970) and extensively modified by us. NLS involves five parameters, namely, geometric time of the
 occultation, velocity component of the moon in the direction of occultation, Signal level, background level and the UD angular diameter. A $\chi^2$
 minimisation technique is followed to obtain the best estimation of the five parameters. The point source Fresnel diffraction pattern modulated by the
 finite telescope aperture, finite optical and electrical bandwidth of the system along with UD angular diameter are used to model the observed light curve.

In case of the InSb photometer (earlier system) faster sampling (upto 1 millisec) is possible for brighter sources (K$\leq$ 3) but the system time response
 has to be taken into account explicitly in the analysis. In case of the subarray operation the sampling time includes integration time, reset time
 and electronics overheads which are kept to a minimum with a small array size. Typically for an integration and reset time of 3 millisec each, the sampling
 time between consecutive subframes is 8.75 millisec. Compared to single element detector sampling is coarser but higher Signal to Noise is possible on
 the light curve. Point source LO light curves obtained by the two modes of operation are shown in Fig.1 along with fitted model curves. Insets in the
 figure show error curves indicating the level of angular resolution achievable. Typically the limiting resolution of the technique with subarrays is about
 3 milliarcsec though single element detector operation can do slightly better for bright sources (K $\leq$ 3).

\begin{figure*}
\centering
\begin{minipage}{170mm}
  \includegraphics[width=17.0 cm,height= 7 cm]{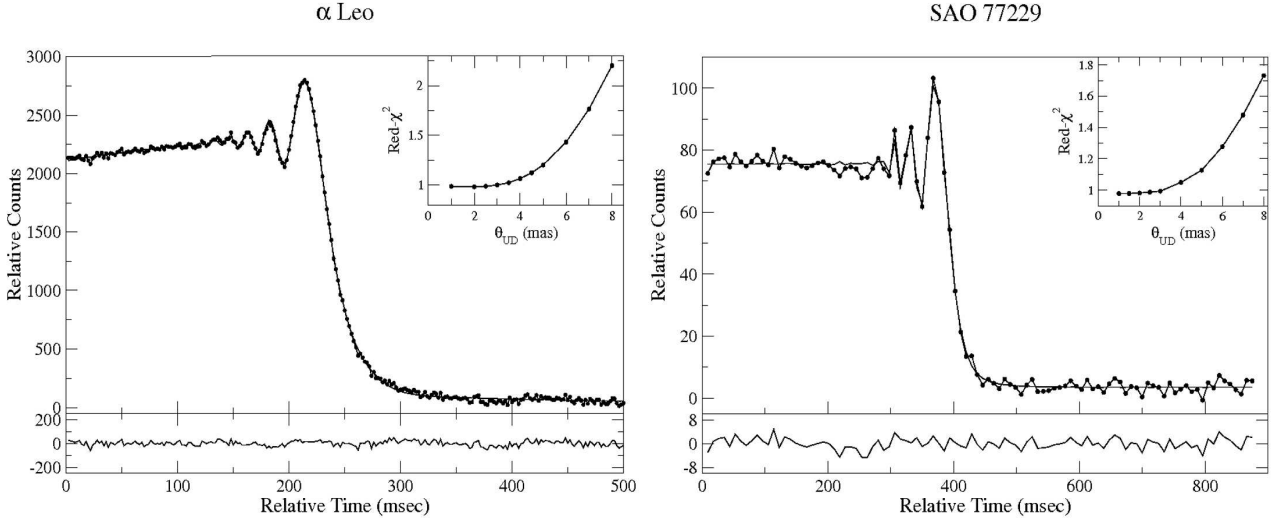}
  \caption{Observed and model fitted light curves of point sources observed by the two different detector systems. The inset shows the error curve
 of the UD angular diameter determination and the lower panel shows the residuals (data-model) of the best fit. The S/N for $\alpha$-Leo (K = 1.6) light
 curve is $\sim$100 and the same for SAO 77229 (K = 3.9) is $\sim$40. The limit of resolution in both the cases is $\sim$3 mas.}
\end{minipage}
\end{figure*}
\section[]{Source details}
The individual source details are listed in Table 2. and discussed below. UD angular diameters derived from our observations are given for each of the
 sources and also listed in Table.3.
\subsection{AW Aur}
AW Aur is a oxygen-rich Mira variable with spectral type ranging from M5 to M9. According to GCVS the reported period is 443.2 days with the epoch of maximum
 light at JD 2453823.0. The V-band magnitude variation is from 10.10 to 17.10 and Spectral type variation from M5 to M9. Using Lomb-Scargle normalized
 periodogram formula (Scargle, 1982) we verified that the maximum power in the light curve is at the period of 445($\pm$10) days in good agreement with the
 reported GCVS value. We adopt this value of periodic variability for AW Aur in this paper.

No Hipparcos parallax measurement is available for this source. However, we can make an estimation of the distance to the source using the
 Period-Luminosity (PL) relation for Galactic Miras as given by Whitelock, Feast \& van Leeuween (2008). We obtain an absolute K-magnitude value -8.18$\pm$0.28
 and the distance to the source 1.02$\pm$0.11 kpc (without extinction correction). This value is in agreement with the value of 1063 pc reported by Le Bertre
 et al. (2003) who also reported a mass loss rate of 3$\times 10^{-7}$ $M_{\odot}$ $yr^{-1}$ for this star. We adopt a value of 1 kpc as distance to the source.

The occultation of AW Aur was recorded close to the maximum (phase 0.95) using the InSb photometer. Following the NLS procedure outlined earlier the
 observed light curve is fitted with the Uniform Disk (UD) model, which is shown in Fig.2 along with the residuals (data-model) in the lower panel. Inset
 in the figure shows the error curve for different UD sizes. The minima of this curve indicates our best estimate for the UD angular diameter. For AW Aur
 we  derive  UD angular size of 4.33$\pm$0.50 milliarcsec. There is no previous measurement of angular size of this source.
\begin{figure*}
\centering
\begin{minipage}{170mm}
  \includegraphics[width=17.0 cm,height= 7 cm]{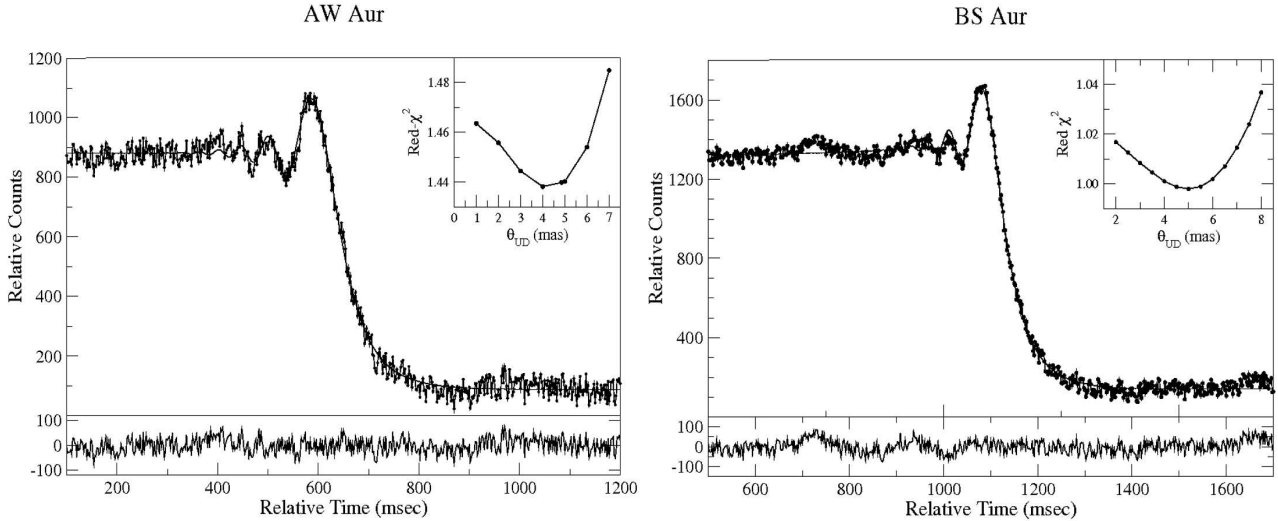}
  \caption{Observed and model fitted light curves of two Mira variables, AW Aur and BS Aur. The inset  shows the error curve of the UD angular diameter
 determination and the lower panel shows the residuals (data-model) of the best fit. Both Miras have been observed in broad $K$-band.}
\end{minipage}
\end{figure*}

\begin{figure*}
\centering
\begin{minipage}{170mm}
  \includegraphics[width=17 cm,height= 14 cm]{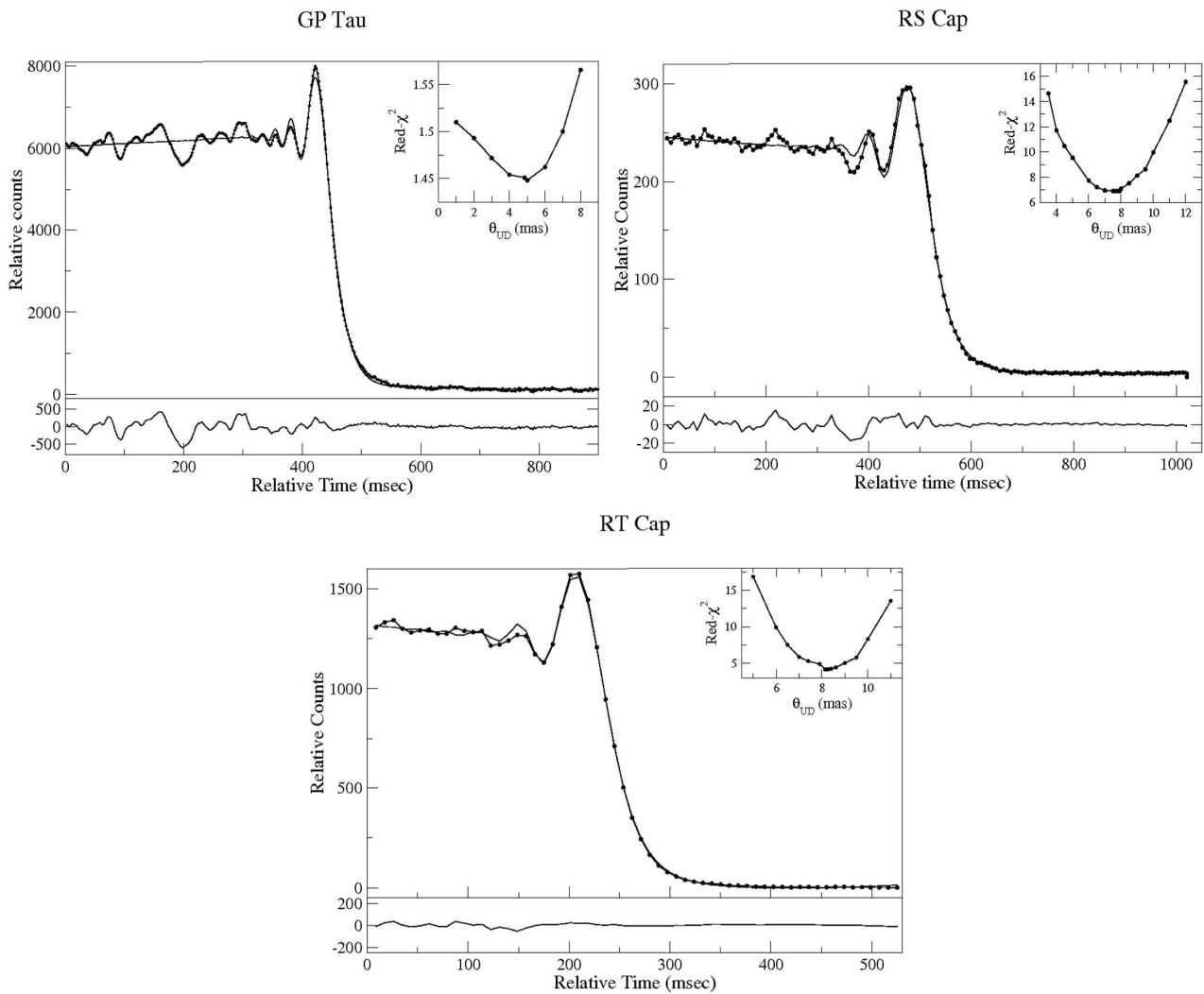}
  \caption{Observed and model fitted light curves of three Semi-regular variables, GP Tau, RS Cap and RT Cap. The inset shows the error curve of the UD
 angular diameter determination and the lower panel shows the residuals (data-model) of the best fit. While GP Tau and RT Cap are observed in the broad
 $K$-band, RS Cap measurements are in a narrow band (2.37 $\mu m$ / 0.1 $\mu m$).}
\end{minipage}
\end{figure*}

\begin{figure*}
\centering
\begin{minipage}{170mm}
  \includegraphics[width=17 cm,height= 7 cm]{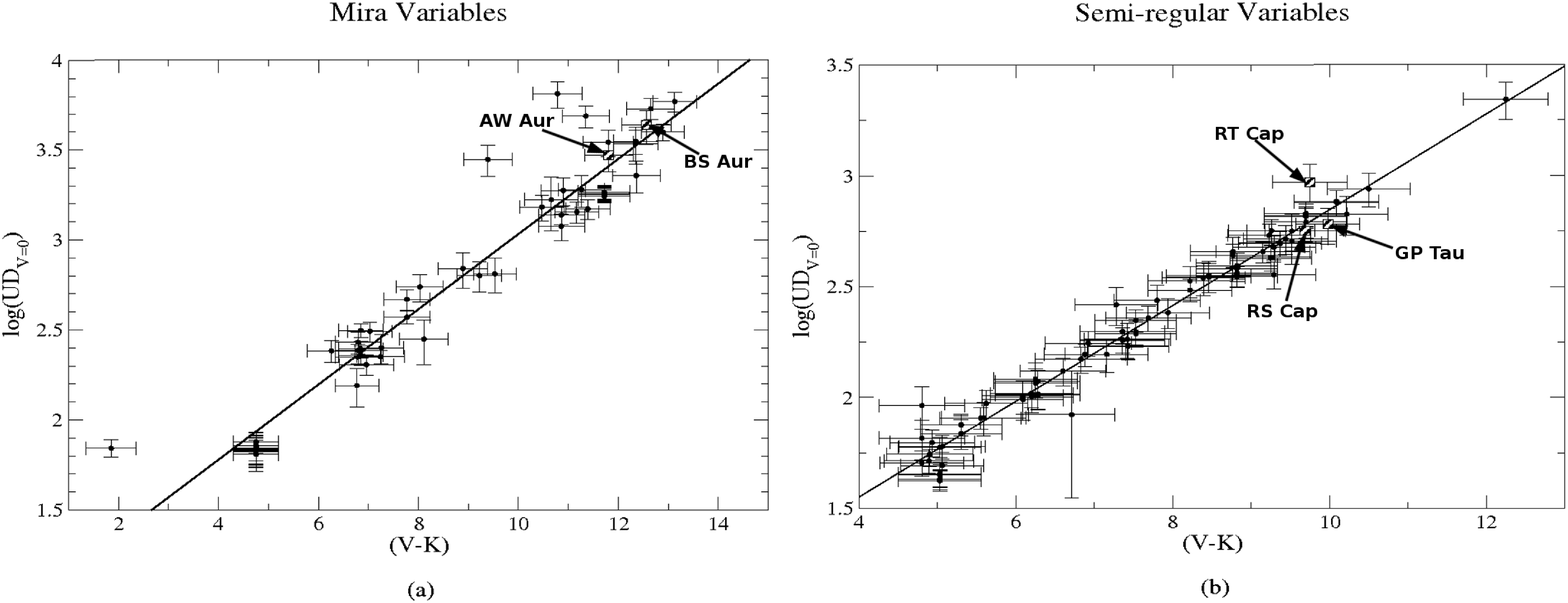}
  \caption{The (V-K) colour vs. zero-magnitude angular diameter of $o$-rich Mira variables (54 measurements) (a) and SRVs (83 measurements) (b). The solid
 line represents the least squares fit to these points. [correlation coefficients are 0.94 (Miras) and 0.98 (SRVs)]. Position of sources in our sample is
 also indicated. Please note that RS Cap measurements are in narrow band filter (2.37$\mu m$/0.1 $\mu m$).}
\end{minipage}
\end{figure*}
\subsection{BS Aur}
BS Aur is a late M-type (M8-M9) oxygen-rich Mira variable with a pulsation period of 466.7 days as reported by GCVS. The V-band magnitude varies from 10.20
 to $>$ 15.00 with the epoch of  maximum light at 2441255.0 JD. Following a similar procedure as in the case of AW Aur we find the maximum power refers
 to the period 480($\pm$5) days which is close to the value reported in the catalog. Using the PL relation as in the case of AW Aur we calculate the
 absolute $K$-mag of the source as -8.27 $\pm$0.20 and estimate a distance of 1.2 kpc (without extinction correction). The LO light curve observed in $NIR$
 $K$-band from Mt. Abu using single channel photometer under good sky conditions is shown in Fig.2. The derived UD angular diameter of the source is
 5.00$\pm$0.70 mas at the photometric phase of 0.83.
\subsection{GP Tau}
 GP Tau is a M7-type giant. According to ASAS catalogue it is a semi-regular (SRb) variable with a period of 109 days and a visual magnitude amplitude of
 0.83 (Pojmanski et al., 2005). It is known from IRAS measurements (Helou \& Walker, 1986) that GP Tau has a thin circumstellar shell (Sloan \& Price (1998).
 $H_2O$ maser has also been reported in the source (Han et.al 1998; Kim J et.al 2010) at a distance of 10-20 stellar radius from the centre.

The only parallax measurement of GP Tau reported has very large error 10.80$\pm$38.30 mas (Hipparcos \& Tycho Cat.). No previous angular size measurement is
 available on this source. Our observed lunar occultation light curve along with its best fit is shown in Fig.3. We obtain the UD angular diameter
 ($\theta _{UD}$) = 4.85$\pm$0.50 mas. We estimate the distance to the source to be $\sim$270 pc, using the absolute magnitude (K=-9.04) for M7 giants as
 reported by Wainscoat et al. (1992)
\subsection{RS Cap}
RS Cap is a late M-type source with a spectral type M6-M7III. It is a semi-regular (SRb) variable. According to the catalog of variable stars in the
 southern hemisphere (Pojmanski et al., 2005) the period of the source is 193 days with a V-amplitude 0.46 mag. Kahane \& Jura (1994) using mm wave
 observations estimated distance to RS Cap of 280 parsecs. Winters et.al (2003) from $CO$ observations report a distance of 277 parsecs and a mass loss
 rate of $\dot{M} = 1.08\times10^{-6} M_{\odot}$ /yr. Earlier angular diameter measurements of RS Cap are also available. Richichi et .al (1992) obtained
 from LO methods the UD angular diameter value of 7.75$\pm$0.67 mas and derived the effective temperature of the source to be $T_e$ = 3560 K. No circumstellar
 shell was detected, though the spectral energy distribution indicates the presence of a weak shell around the source with less than 1\% strength of the
 stellar signal. Later Dyck, van Belle \& Thompson (1998) reported from interferometric observations at 2.2 $\mu m$ an angular diameter of  7.0$\pm$0.8 mas.

We obtain the best fit UD angular diameter of 7.70$\pm$0.50 mas which is consistent with earlier measurements. We adopt a distance to the source of 280 pc.
\subsection{RT Cap}
According to  Bergeat \& Chevallier (2005) RT Cap is a carbon-rich, non-Mira giant with photospheric carbon to oxygen ratio (C/O) 1.10. The periods derived
 from $V$-band light curves are 393 days (GCVS) and 389 days (Pojmanski et. al. 2005). The distance to the source is estimated to be 560 pc using apparent
 and absolute bolometric magnitudes 3.80 and -4.95 respectively (Bergeat, Knapik \& Rutily, 2002). They also derived an effective temperature $T_e$ = 2480 K and a
 mass-loss rate of $2.3\times10^{-7} M_\odot yr^{-1}$.
 
There are two high angular diameter measurements reported for this source earlier. One earlier measurement by Schmidtke et al. (1986) using the Lunar
 Occultation technique in narrow $K$-band (2.173 $\mu m$/BW 0.032 $\mu m$) yielded a value of $\theta_{UD}$ = 7.72$\pm$0.16 mas at a photometric phase of
 0.98. Another value reported using long-baseline interferometry in broad $K$-band (van Belle et al., 2000) is $\theta_{UD}$ = 8.18$\pm$0.21 mas.

From our observed light curve (Fig.3) we derive a UD value of  $\theta_{UD}$ = 8.14$\pm$0.50 mas at a photometric phase of 0.72.
\begin{figure*}
  \includegraphics[width=17 cm,height= 7 cm]{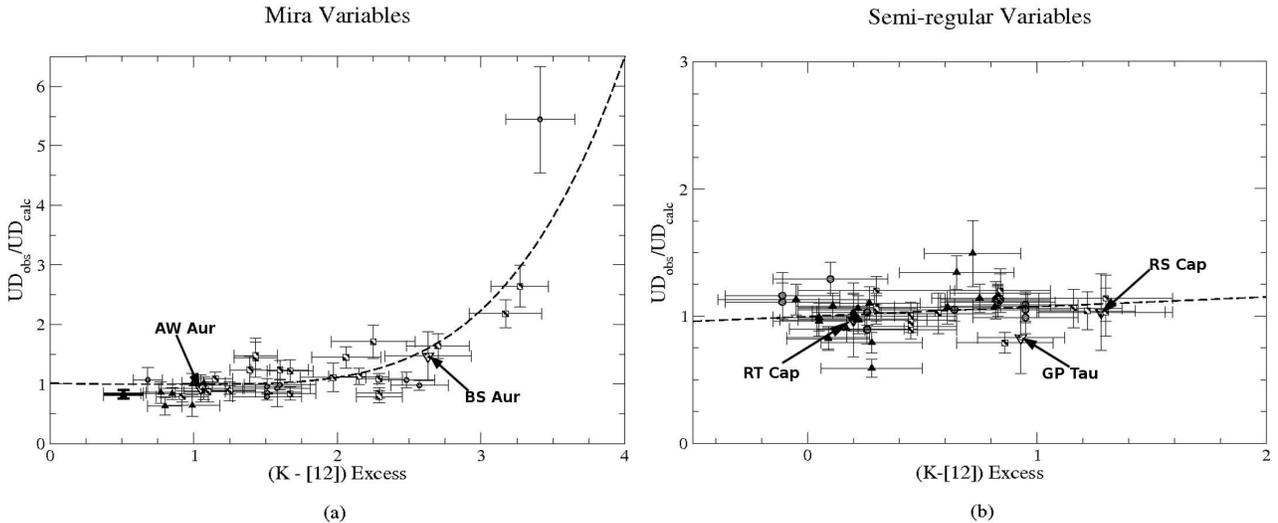}
  \caption{The ($K-[12]$) colour excess vs. the ratio of observed and calculated angular diameters for both Mira variables (a) and Semi-regular variables
 (b) as described in the text. The dotted lines show a polynomial through the points to indicate the rising trend. Our observed sources are also plotted
 and labelled. Please note that RS Cap measurements are in narrow band filter (2.37$\mu m$/0.1 $\mu m$).}
\end{figure*}
\section[]{Results and Discussions}
Uniform disk (UD) Angular diameters of the five sources derived from our observations and analysis are listed in second column of Table.3.
\subsection{UD Angular Diameter predictions using ($V-K$) colour}
We have first made an attempt to compare our results with the predictions of angular sizes generated by us following the approximate methods devised by Di
 Benedetto (1993) and van Belle (1999). The methods use the observed $K$ and $V$ broad band photometry to predict a zero magnitude ($V=0$) angular size
 using $(V-K)$ values through a calibration. The zero magnitude angular sizes are then scaled to the apparent angular sizes using $V$-band photometry. We
 have generated a new calibration using the 54 measured $K$-band angular diameter determinations of oxygen-rich Miras available in the literature (Richichi
 et al., 2005), scaled to the zero-magnitude angular diameters, and plotted against their respective $(V-K)$ values (Fig.4). We have also derived a similar
 relationship for SRVs using 83 sources from the same catalog (Fig.4). A good correlation is obtained in both cases (correlation coefficients of 0.94 (Miras)
 and 0.98 (SRVs)). It must be pointed out that the errors involved in the angular diameter predictions by these methods is in the range 20 to 25 $\%$ due to
 difficulties of obtaining contemporaneous photometry. Nevertheless the method outlined appears to have a good predictive value for angular diameters using
 only ($V-K$) colour for both Miras and SRVs.

The predicted UD angular diameters for our 5 sources are listed in third column of Table.3. It is seen that there is a good agreement in general between
 our measured values and predictions. However, in case of RT Cap the prediction gives a lower value. It must be pointed out that RT Cap is a carbon rich
 semi-regular variable, and so it cannot be readily compared with generally oxygen rich SRVs used in the calibration curve.
\subsection{Enhancement of Angular diameter due to Mass loss/shell effects in Miras}
In the case of the two Miras in our sample we have also investigated the possibility of high mass loss rates and presence of a circumstellar shell affecting
 the angular size measurements. This aspect of enhancement of UD angular size due to mass loss was discussed earlier by van Belle, Thompson \& Creech-Eakman
 (2002). These authors had derived, for a sample of Miras, the ratio of linear radius (obtained from angular size measurements and distance estimates) to the
 theoretical (Rosseland) radius assuming that Miras are all fundamental mode pulsators. A plot of this ratio as a function of $(K-[12])$ colour excess which
 is indicative of mass loss showed that Miras with higher colour excess were systematically 120 $R_{\odot}$ larger than their counterparts with lower colour
 excess, independent of the periods.

We have adopted a slightly different approach using a sample of 43 oxygen rich Miras with spectral types later than M5. In order to avoid the large errors
 involved in distance measurements we plot the ratio of observed and calculated angular diameters against $(K-[12])$ colour excess. The $K$-band magnitudes
 and 12 $\mu m$ fluxes are collected from 2MASS catalog (Cutri et al., 2003) and IRAS catalog (Helou \& Walker, 1986) respectively. The calculated angular
 diameter is derived from bolometric flux and effective temperature, and is also independent of distance. The bolometric fluxes are calculated using the mean
 relation between bolometric flux, $K$-band flux and ($V-K$) colour as reported by Dyck, Lockwood \& Capps (1974) and the corresponding effective temperatures
 are collected from Alvarez \& Mennessier (1997). The ratio of reported UD angular diameters (Richichi et al., 2005) and calculated angular diameters
 ($UD_{calc}$) are then plotted against the $(K-[12])$ colour excess (Fig.5a). The plot shows that the ratio remains close to unity upto a colour excess
 $\sim 2.5$, and then increases sharply. For a colour excess of $\sim$3 the measured UD diameter is almost twice the calculated value. The measured UD
 diameter of AW Aur is in good agreement with the calculated value showing a ratio $\sim$1. However the observed value of BS Aur is almost 1.4 times the
 calculated value (Table.3, column 4). We have also plotted the positions of our sources (AW Aur and BS Aur) in Fig.5a which suggests that BS Aur has a high
 mass loss rate and may harbour a shell. This is also borne out by the IRAS LRS characterization of these two stars. According to this characterization, AW
 Aur (LRS Char 15) does not have any detectable circumstellar shell but BS Aur (LRS Char 28) has a thin $o$-rich shell around it. 

Following the above procedure we have also carried out a similar investigation for SRVs using a sample of 52 $K$-band angular diameter measurements (later
 than M5) taken from Charm2 catalog (Richichi et al., 2005). The SRVs in the sample have a ($K-[12]$) colour excess less than 1.5 (unlike Miras) and the
 ratio close to unity (Fig.5b). It appears unlikely that SRVs may exhibit enhancement in their angular diameter.

Using the available distance estimates of each of these sources given in section 3 we have derived, using our measured UD angular diameters the linear radii
 of the sources corrected for enhancement. With a reduction of 40\% of the enhanced size of BS Aur we find that both Miras have a similar radii (Table.3,
 column 5). However, it may be pointed out that due to large unknown errors in distance to these sources the absolute errors involved in linear radii could be
 much higher. Hence it is difficult to draw any conclusion regarding the mode of pulsation of these two Miras.

It is speculated that SRVs can pulsate in a number of modes and that too often simultaneously (Lattanzio \& Wood, 2003). However, in the absence of reliable
 distance measurements it is difficult to draw definitive conclusions from our angular diameter measurements.
\begin{table}
 \centering
  \caption{Results.}
  \begin{tabular}{@{}lcccccccccc@{}}
  \hline
  Source &Obs. $\theta_{UD}$&Pred. $\theta_{UD}$&    Ratio    &Linear Radius\\
         &     (mas)        &      (mas)        &  $\frac{UD_{obs}}{UD_{calc}}$ & (Corrected) \\
         &                  &                   &             & (R$_\odot$) \\
  \hline 
  AW Aur &  4.33$\pm$0.50   &   4.0$\pm$1.0     &  1.1$\pm$0.1& 440$\pm$100 \\
  BS Aur &  5.00$\pm$0.70   &   4.5$\pm$1.2     &  1.4$\pm$0.2& 470$\pm$110 \\
  \hline
  GP Tau &  4.85$\pm$0.50   &   6.1$\pm$1.5     &  0.8$\pm$0.1& 175$\pm$ 40 \\
  RS Cap*&  7.70$\pm$0.50   &   8.3$\pm$2.1     &  1.0$\pm$0.1& 230$\pm$ 40 \\
  RT Cap &  8.14$\pm$0.30   &   5.4$\pm$1.3     &  1.0$\pm$0.1& 490$\pm$ 70 \\
\hline 
\end{tabular}
\begin{flushleft}
* RS Cap has been observed in narrow $CO$-band filter (2.37$\mu m$ /0.1 $\mu m$).
\end{flushleft}
\end{table}

\section{Conclusions}
$K$-band uniform disk angular diameters of two Mira variables and three Semi-regular variables (SRVs) are reported. For the two Miras and one SRV (GP Tau)
 these are the first time angular diameter measurements. For the other two SRVs our values are in good agreement with those reported earlier. Two separate
 comparative studies have been made to examine our measured values with predictions. One of the methods involves a separate calibrations for Miras and SRVs
 with previously reported $K$-band diameters and ($V-K$) colours. In this case we find a good agreement between measurements and predictions except for one
 SRV (RT Cap) which is a carbon star. We have also investigated the enhancement of measured UD angular diameter due to heavy mass loss and the presence of a
 circumstellar shell as indicated by ($K-[12]$) colour excess. We find AW Aur is unlikely to harbour a shell. But for the other Mira in our sample (BS Aur)
 UD angular diameter measured appears enhanced by nearly 40\% compared to the expected value due to presence of a circumstellar shell arising out of mass loss.
\section*{Acknowledgments}
This work was supported by Dept of Space, Govt of India. This research made use of the SIMBAD data base operated at the CDS, Starsbourg, France and catalogues
 associated with it. We thank the referee for his valuable comments.

\label{lastpage}

\end{document}